\documentclass[11pt,titlepage]{article}
\usepackage{setspace} 
\usepackage{dcolumn,color,helvet,times}
\usepackage{amsmath,lastpage,bm,textcomp}
\usepackage{subfigure, amssymb, wasysym, chemarr}
\usepackage[round,numbers,sort&compress]{natbib}
\usepackage{multicol}
\usepackage{graphicx}

\newcommand{\bN}{\mbox{\boldmath $N$}}

\newcommand{\bD}{\mbox{\boldmath $D$}}
 
\newcommand{\bs}{\mbox{\boldmath $s$}}

\newcommand{\blambda}{\mbox{\boldmath $\lambda$}}

\newcommand{\bJ}{\mbox{\boldmath $J$}}
\newcommand{\bp}{\mbox{\boldmath $p$}}
\newcommand{\bx}{\mbox{\boldmath $x$}}
\newcommand{\by}{\mbox{\boldmath $y$}}

\newcommand{\bsigma}{\mbox{\boldmath $\sigma$}}

\newcommand{\bv}{\mbox{\boldmath $v$}}

\title{Stochastic Control Analysis for Biochemical Reaction Systems}        

\author{Kyung~Hyuk~Kim \thanks{
	   Kyung Hyuk Kim.  Address: 
           Department of Bioengineering,
	   University of Washington,
	   William H. Foege Building
	   Box 355061 
	   Seattle, WA 98195-5061, U.S.A.,
	   Tel.:~(206)543-4291} \\
	Department of Bioengineering, \\
        University of Washington, Seattle, WA 98195-5061, U.S.A.
	\and Herbert~M.~Sauro \\
	Department of Bioengineering, \\
        University of Washington, Seattle, WA 98195-5061, U.S.A.}

\date{}

\begin{document}

\maketitle

\abstract{
We investigate how stochastic reaction processes are affected by external perturbations.  We describe an extension of the deterministic metabolic control analysis (MCA) to the stochastic regime.  We introduce stochastic sensitivities for mean and covariance values of reactant concentrations and reaction fluxes and show that there exist MCA-like summation theorems among these sensitivities.  The summation theorems for flux variances and the control distribution of the flux variances is shown to depend on the size of the measurement time window ($\epsilon$) within which reaction events are counted for measuring a single flux.  It is found that the degree of the $\epsilon$-dependency can become significant for processes involving multi-time-scale dynamics and is estimated by introducing a new measure of time scale separation.  This $\epsilon$-dependency is shown to be closely related to the power-law scaling observed in flux fluctuations in various complex networks.   We also propose a systematic way to  control fluctuations of reactant concentrations while minimizing changes in mean concentration levels by applying the stochastic sensitivities.  
}

\clearpage

\section*{Introduction}
Metabolic control analysis (MCA)~\cite{Fell1992, Kacser1995,Fell1996}  and the closely related biochemical systems theory \cite{Savageau1976} have greatly enhanced our ability to understand the dynamics of cellular networks. However, these approaches are based on a deterministic picture of cellular processes and in recent years it has become clear that many networks, such as gene regulatory networks, operate with a significant degree of stochasticity~\cite{Arkin1998,  Elowitz2002, Ozbudak2002, Rosenfeld2005, Austin2006, Elf2007}. In these situations a deterministic formalism is inadequate \cite{Rao2002,Raser2004,Kaern2005,Shahrezaei2008a}.  In this paper we begin the process of developing a new analysis method of control on stochastic dynamics by extending MCA to the stochastic regime. We call the extension stochastic control analysis (SCA).

MCA is an analysis of sensitivities which quantifies how much system variables change in response of the perturbations in system parameters.  To extend the MCA to the stochastic regime, we need to introduce sensitivity measures for stochastic system variables. There have been a wide variety of efforts in recent years to introduce and investigate sensitivity measures for stochastic systems related to  \emph{mean} levels of concentrations and fluxes in stochastic reaction systems~\cite{Paulsson2000, Rao2002, Thattai2002, Elf2003, Pedraza2005, Hooshangi2005, Qian2006}. More pertinent to this paper is the work by Andrea Rocco who investigated the MCA summation and connectivity theorems related to the most-probable concentration values and their corresponding reaction rates \cite{Rocco2009}. However, the sensitivities for noise characteristics (variance and covaraince) were not investigated and the summation theorems related to the noise properties were not discussed. Thus, a systematic MCA-like approach for controlling noise has not been made.

In this paper we will focus on the control coefficients~\cite{Fell1992, Kacser1995,Fell1996}, for \emph{variances} and \emph{covariances} of concentrations and fluxes.  The control coefficients quantify the global responses due to (static) perturbations in the system parameters. We also introduce sensitivities for the mean levels of concentrations and fluxes, which are closely related to the MCA control coefficients.  We obtain MCA-like summation theorems for the stochastic sensitivity measures. In a similar way to the deterministic MCA theorems, the SCA theorems imply that control is distributed over a reaction system.  

The summation theorems for flux variances show very interesting properties:  
Flux is measured by counting the number of reaction events within a given time window $\epsilon$. From this we show that the sum value can be highly dependent on the measurement time window ($\epsilon$).  This in turn implies that the control of flux variances can be sensitive to the value of $\epsilon$.  We present a case where a control distribution over a reaction system changes significantly with the value of $\epsilon$ and thus the system can show an increase or decrease in the flux variances depending on the value of $\epsilon$. The degree of such $\epsilon$-dependency of the flux variance control is closely related to how both the time scales of the fast and slow fluctuations are separated.  Such a separation can be quantified by introducing a new time scale separation measure which can be estimated from the temporal sequences of reaction events.

The summation theorems for flux variances also show a close connection to the scaling relationship between flux variances and their mean values recently observed in various complex networks: the Internet, microprocessor logic networks, the World Wide Web, highway systems, river networks, and stock markets \cite{Menezes2004, Menezes2004b,Eisler2005, Eisler2005b, Meloni2008}.  In these systems, fluxes were defined as the number of packets processed in network routers in the Internet, activity of connections between logic gates in the microprocessors, the number of visits on sites in the World Wide Web, the number of cars of traffic at different locations in the highway systems, stream flows in the river networks, and the traded values of stocks in the stock market.  It was investigated how the standard deviation ($\sigma$) of the flux is related to the mean value of the flux ($\langle f \rangle$): $\sigma \sim \langle f \rangle^\alpha$.  de Menezes and Barab{\'a}si claimed that the Internet and microprocessor logic networks belong to a universality class characterized by an exponent value of $\alpha = 0.5$, and the World Wide Web, highway systems, and river networks to that of $\alpha = 1$ \cite{Menezes2004}.  However, stock markets such as NYSE and NASDAQ show non-universal values of $\alpha$ \cite{Eisler2005, Eisler2005b}.  Meloni et al. \cite{Meloni2008} proposed a model of random diffusion to show how the value of the exponent can crossover from 0.5 to 1 depending on the number of links connected to a node, the strength of external noise and the time measurement window size $\epsilon$. In this paper, we show a connection between the summation theorems for flux variances and the scaling crossover phenomena. We briefly discuss that the exponent crossover can take two different forms depending on the time window size $\epsilon$ relative to the correlation time of the external noise.

As an application of the stochastic control coefficients, we provide a systematic non-local method for the control of noise levels in concentrations, while minimizing changes in mean concentration values.  Such orthogonal control is performed first by estimating control coefficients for mean values and coefficients of variation (CVs) of concentrations under the linear noise approximation \cite{vankampen2007}.  From the estimate of the control coefficients, we find the direction of the parameter perturbations leading to a sensitive response of change in the CVs of the concentrations while minimizing the change in their mean values and this enables us to identify which sets of parameters need to be controlled by how much in a relative sense.  We apply this orthogonal control in a negative feedback system under external noise and successfully reduce the concentration noise level, with a negligible change in the concentration mean value.

\section*{Model Systems and Definitions of Control Coefficients}

The model system we will consider is a chemical reaction system described by the chemical master equation \cite{Mcquarrie1967, Gillespie1992}, i.e., we assume the system is spatially homogeneous (uniform concentrations throughout the time evolution of the system).  We assume that it can be described by $L$ kinds of reaction rates for $M$ reactants. The system is composed of the external and internal processes. The external process is modeled by allowing one of the species (denoted by either $S_e$ or $S_1$) to fluctuate slowly and independently, compared to the rest.  $S_e$ is considered a source of external noise.  The internal system, composed of all other species,  is affected by the external noise and also by internal noise generated from the internal reactions.  

To estimate how a system responds under parameter  perturbations at the stationary state, we introduce sensitivity measures called control coefficients.  The system variables ($y$) of interest can be either mean values or coefficients of variation/covariation (CV/CCV) of concentrations and reaction fluxes. CV is variance divided by the mean square and CCV is the covariance (between two variables) divided by the product of their mean values.   We define the control coefficients for these variables as
\[
C^y_p =\frac{p}{y}\frac{d y}{dp}=  \frac{d \log y }{d\log p},
\]
which indicates the relative change in $y$ due to a given relative change in a parameter $p$.  The change in $y$ is from one stationary state to another corresponding to before and after the perturbation, respectively.   We note that control coefficients for different system variables -- most-probable concentrations (not mean concentrations) -- have been investigated in the framework of MCA, but sensitivities related to fluctuation properties have not \cite{Rocco2009}.  The parameter $p$ will be called here a control parameter, which is not affected by the system's reactions.  We restrict the set of the control parameters ($\bp = (p_1, \cdots, p_L$)) to be the proportionality constants of reaction rates. E.g., for a reaction rate $v= \frac{p \, s}{K_M + s}$ with $s$ concentration and $K_M$ a Michaelis constant, $p$ is a control parameter but $K_M$ is not.  The total enzyme concentration that catalyses a reaction is one such parameter.

\section*{SCA: Summation Theorems for Control Coefficients} 
We have found that there exist MCA-like summation theorems among the proposed stochastic sensitivities, which are valid under \emph{any} strength of noise and \emph{finite} perturbations of parameters $\bp$.   The existence of these theorems is rooted in the fact that the stochastic measures satisfy certain scaling properties under a specific kind of scale change in time and control parameters. 

\subsection*{Summation theorems for concentrations}
We note that all reaction propensity functions $v_i$ are proportional to control parameter $p_i$: $v_i(s, \alpha \bp) = \alpha v_i(s, \bp)$.   Let us scale all control parameters by a fixed proportion $\alpha$.    The simultaneous change in all propensity functions can be interpreted as a change in the time scale in the amount of $1/\alpha$  because the propensity functions are inversely proportional to time.  
Mean levels, CVs and CCVs of concentrations are time independent variables at stationary states.   This means that these quantities remain the same under the parameter change \cite{Giersch1988b}.   We can summarize these arguments with the following equation (refer to Table~\ref{table-notation} for notation).  The change in a concentration mean level is expressed as:
\[
\delta \langle s_j\rangle = \sum_i C^{\langle s_j\rangle}_{p_i}
\frac{\delta p_i}{p_i} = \alpha\sum_i C^{\langle
s_j\rangle}_{p_i},
\] 
for all $j=0,\cdots, M$.  Since $\delta \langle s_j\rangle =0$, we derive 
\begin{equation}
\sum_{i=1}^{L} C^{\langle s_j \rangle}_{p_i} = 0,
\label{eqn:sum-s-mean}
\end{equation}
for all species $j$.    The same argument can be applied for the concentration CVs and CCVs. 
\begin{equation}
\sum_{i=1}^{L} C^{V_{jk}^s}_{p_i} = 0,
\label{eqn:sum-s-CV}
\end{equation}
for all species $j$ and $k$.

\subsection*{Summation theorems for fluxes}
To derive the summation theorems for mean fluxes, we consider again the parameter scale change. Under this change, mean propensity functions will scale by $\alpha$.  Since the mean propensity function $\langle v_l \rangle$ is equal to the mean fluxes $\langle J_l \rangle$, the mean fluxes will also scale by $\alpha$ \cite{Giersch1988b}.  Since the scale change in the mean flux can be expressed as:
\[
\frac{\delta \langle J_l \rangle}{\langle J_l \rangle} = \sum_{i=1}^M C^{ \langle J_l \rangle }_{p_i} \frac{\delta p_i}{p_i} =  \frac{\delta p}{p}\sum_i C^{\langle J_l \rangle}_{p_i} = \alpha \sum_i C^{\langle J_l \rangle}_{p_i},
\]
we obtain summation theorems for mean flux control coefficients:
\begin{equation}
\sum_{i=1}^{L} C^{\langle J_{j} \rangle}_{p_i} = 1.
\label{eqn:sum-mean-J}
\end{equation}

We will also derive summation theorems for flux CVs and CCVs. However before we derive them, it is important to clarify the difference between  a propensity function, a reaction rate, and a reaction flux.  All of them are stochastic variables.  The reaction flux $J$ is measured by counting the number of reaction events within a time window $\epsilon$:
\[
J_i = \frac{ \mbox{\begin{tabular}{c} Number of events of ($i$-th kind) reaction   \\ occurred during $\epsilon$ \end{tabular} } }{\epsilon}.
\]
The propensity function is a mathematical function previously denoted by $v$.  The mean values of both $v$ and $J$ are equal.    The fluctuation strengths of each can however be different, because the variances of $J$ are dependent on $\epsilon$ (as will be discussed later), while those of $v$ are not.  We express the CV/CCV of $J$ ($V^J$) as a function of $\epsilon$: $V^J(\epsilon, \bp)$.   We use the  term, reaction rate, as either the flux or propensity function, depending on context.  

Now we will derive the summation theorems for flux CVs. The first thing to note is that flux CVs are unitless in time.  The flux CVs obtained by scaling all parameters by $\alpha$ is the same as those obtained by scaling the time by $1/\alpha$:
\[
V^{J}\left(\epsilon, \alpha \bp\right) = V^{J}\left(\alpha \epsilon, \bp\right).
\]
This can be expressed in terms of control coefficients as follows:
\begin{equation}
\sum_{i=1}^{L} C^{{V_{jk}^{J}}}_{p_i} = \frac{\partial \log V_{jk}^{J}}{\partial  \log \epsilon},
\label{eqn:theorem2}
\end{equation}
for all reactions $j,k$.
This equation means that the sum value is \emph{equal} to the slope of a log-log plot of flux CV and CCV vs. $\epsilon$. Since the flux CV and CCV depend on $\epsilon$, the sum value can also depend on $\epsilon$.

\subsection*{Summation theorems for flux CVs vs. multi-time-scale dynamics}
 In this section, we investigate how the sum value of Eq.~\ref{eqn:theorem2} changes with $\epsilon$.  We have found an interesting fact that the sum value can vary significantly with the change in $\epsilon$ when the system shows wide distributions of reaction time scales.    

\begin{figure}
\hspace{3cm}
\includegraphics[width=6cm, angle = 0]{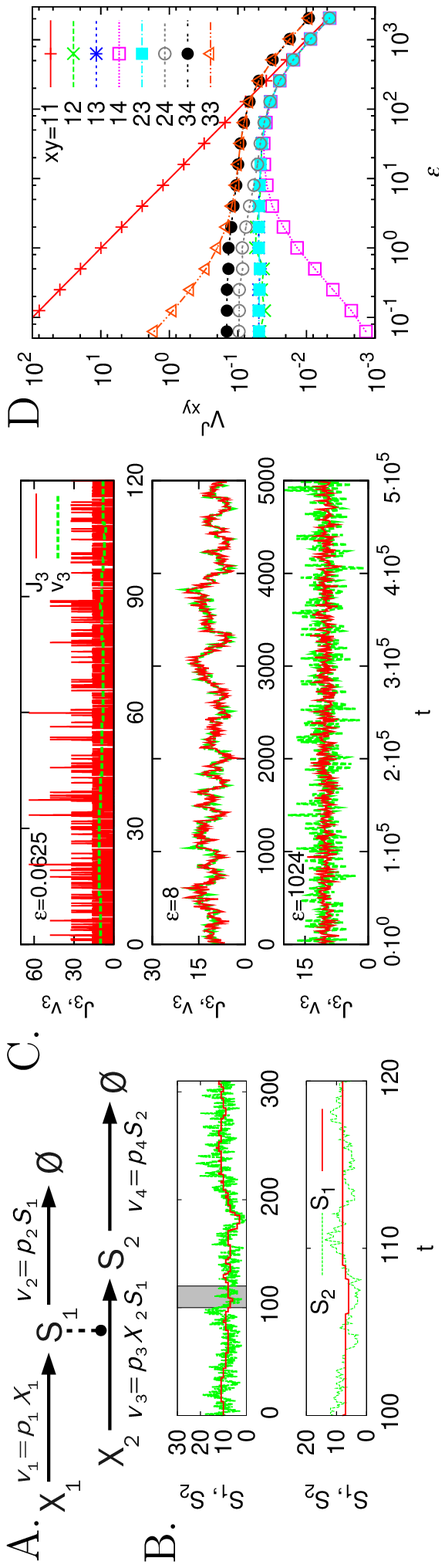}
\caption{}
\label{fig:2step}
\end{figure}

We briefly discuss the mechanisms for leading to the sum value change by considering a simple reaction system: a two-step reaction cascade as shown in Fig.~\ref{fig:2step}A.  $S_1$ is created with a rate $v_1$ and degrades with a rate $v_2$. $S_1$  enhances the conversion of $X_2$ to $S_2$.     We assume that the creation and degradation processes of $S_1$ are much slower than those of $S_2$.  $S_1$ is the source of external noise. The reaction process involving $S_2$ is considered an internal system.    The time evolution trajectory of $S_2$ shows a mixture of two different kinds of noise (slow and fast) as shown in Fig.~\ref{fig:2step}B \cite{Menezes2004, Menezes2004b, Meloni2008}. In the time resolution of $\epsilon \sim 0.1$, external noise is negligible while the internal noise (caused by reaction events of $v_3$ and $v_4$) is dominant. As $\epsilon$ increases, the external noise becomes more dominant while the internal noise becomes averaged out. The creation flux of $S_2$ shows this tendency clearly as shown in Fig.~\ref{fig:2step}C.

\begin{figure}
\includegraphics[scale = 1]{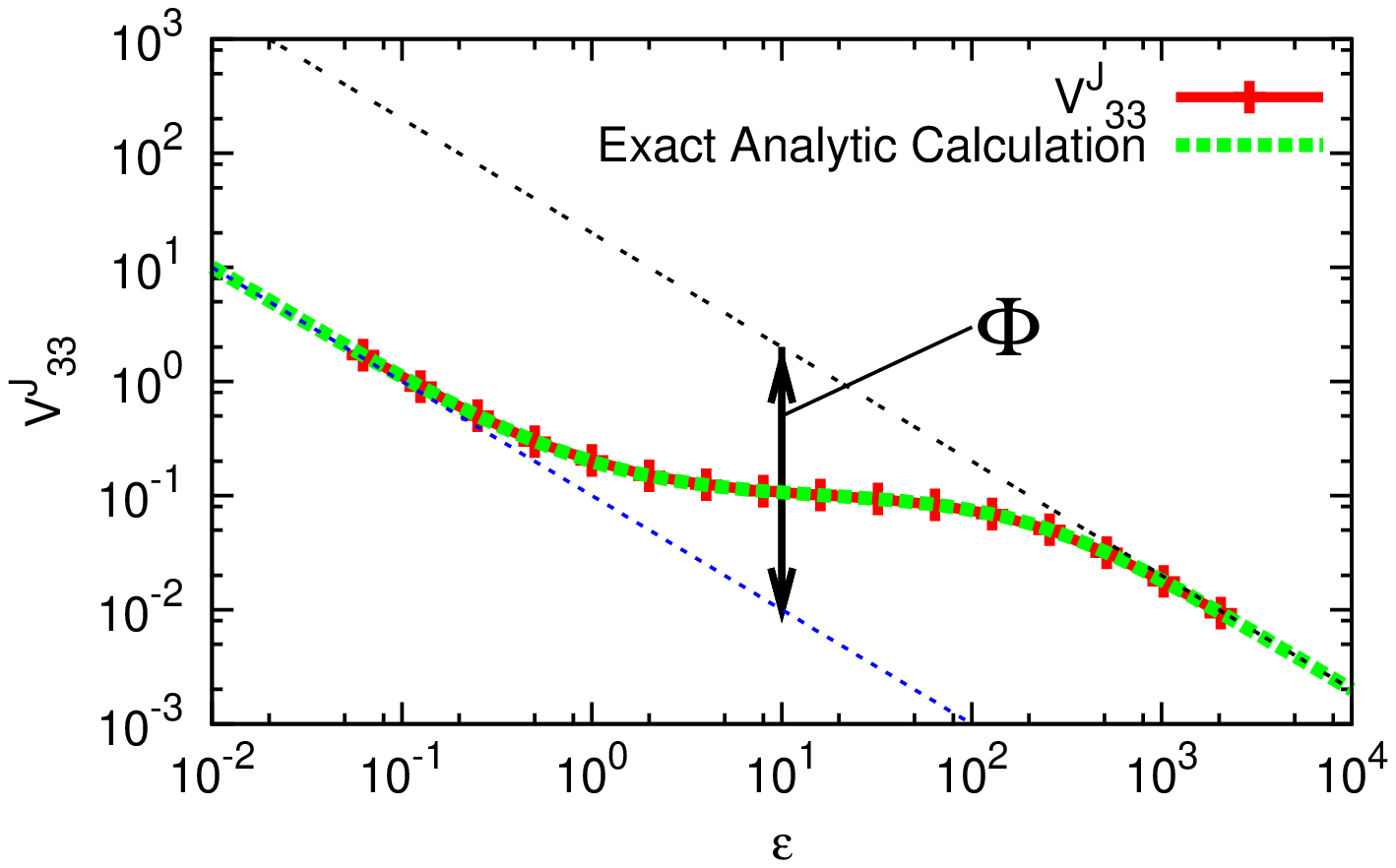}
\caption{}
\label{fig:2step-b}
\end{figure}

We have plotted all internal and external flux CVs and CCVs vs. $\epsilon$ (Fig.~\ref{fig:2step}D). First, we discuss the flux CVs: $V^J_{ii}$. For the fluxes corresponding to the fast reactions ($i=3,4$), a plateau region appears (slope $\sim 0$, i.e., the sum value $\sim 0$) and for the fluxes corresponding to the slow reactions ($i=1,2$), they don't.  The plateau region appears due to the fact that the internal noise becomes sufficiently averaged out at the time scale of $\epsilon \sim 1/p_3=1$ and the external noise becomes dominant for all values of $\epsilon \gtrsim 1/p_3$. $J_3$ can be approximated to be $v_3$ for $\epsilon \sim 10$ (Fig.~\ref{fig:2step}C): $J_3 \simeq v_3=p_3 S_1$.  The flux CV can be expressed as $V^J_{33} \simeq  V^S_{11} = \frac{1}{\langle S_1 \rangle}=0.1$. This is the value of the flux CV at the plateau region.  This is because the external noise has the correlation time ($1/p_2$) and the approximate equality $J_3 \simeq v_3$ persists until the external noise is correlated in time, up to $\epsilon \sim 1/p_2=100$. Therefore, the slope of the log-log plot of $V^J$ vs. $\epsilon$ becomes close to zero (Fig.~\ref{fig:2step-b}), which means that the sum value of the flux CV is also close to zero.

For $\epsilon \ll \tau (\equiv 1/p_2)$, $S_1$ does not fluctuate compared to $S_2$. $S_2$ can be considered to be created from a constant source.  The probability $P(n;\epsilon)$ of having the number $n$ of events of reaction $v_3$ during time $\epsilon$ satisfies a Poisson distribution:
\[
P(n;\epsilon) = e^{-v_3 \epsilon} \frac{(v_3 \epsilon)^n}{n!}.
\]
The flux CV becomes inversely proportional to $\epsilon$ (Fig.~\ref{fig:2step-b}):
\[
V^J_{33} = \frac{\langle J_3^2 \rangle- \langle J_3 \rangle^2}{\langle J_3 \rangle^2} =\frac{\langle n^2 \rangle - \langle n \rangle^2 }{\langle n \rangle^2} = \frac{1}{\langle n \rangle} = \frac{1}{\epsilon \langle J_3 \rangle}. 
\]
Thus, the sum value of the flux CV control coefficients is -1.

For $\epsilon \gg \tau $, the external noise becomes uncorrelated in time at this time scale. Thus, the flux $J_3$ measured by using this $\epsilon$ value will be uncorrelated (statistically independent) in time.  We denote the minimum of such a value of $\epsilon$ by $\epsilon_{ind}$.   For the value of $\epsilon \gg \epsilon_{ind}$, the flux estimate $J^\epsilon$ can be considered an average of independent samples of $J^{\epsilon_{ind}}$ with a sample size $\epsilon/\epsilon_{ind}$.   Therefore, $V^{J^\epsilon} = V^{J^{\epsilon_{ind}}}\frac{1}{\epsilon/\epsilon_{ind}} \propto \frac{1}{\epsilon}$.  (From Fig.~\ref{fig:2step-b}, $\epsilon_{ind}$ is $\sim 10^3$.)
This explains intuitively why the flux CV scales as $1/\epsilon$ for large $\epsilon$ values (Fig.~\ref{fig:2step-b}).   
Therefore, the sum value of the flux CV control coefficients is -1. 

For each different pair of fluxes, the asymptotic form of its coefficient of covariance for $\epsilon \ll \tau$, is different: either a plateau or a straight line proportional to $\epsilon$.   A detailed discussion on this is provided in the Supporting Material. 

The arguments presented above can be generalized for a typical reaction systems showing flux fluctuations with two different time scale dynamics (refer to the Supporting Material).   A plateau region (for intermediate $\epsilon$) and two regions of -1 slope (for very small $\epsilon$ and large $\epsilon$) can appear typically for CVs of such fluctuations.

\subsection*{Estimation of time scale separation}

As presented previously, the plateau region in Fig.~\ref{fig:2step-b} appears due to the time scale separation between fast and slow system dynamics.  If the separation is not wide enough, the plateau region can be tilted.  In this case, the sum value of the flux CV control coefficients will deviate from zero in the region of the plateau.  To identify  such deviations, we propose a time-scale separation measure. 

The separation measure ($\Phi$) quantifies the vertical distance between the two asymptotic linear lines for the log-log plot of flux CV vs. $\epsilon$ corresponding to $\epsilon \rightarrow 0$ and $\infty$ as shown in Fig.~\ref{fig:2step-b}. The larger the measure $\Phi$, the wider the plateau region and the smaller its slope, i.e., the sum value of flux CV control coefficients becomes closer to zero.  Consider a reaction step with its propensity function given by $v(s_e)$, explicitly showing a dependency on the external species $S_e$ acting as a source of slow noise.  If events in the above reaction $v(s_e)$ do not cause large fluctuations in substrate concentrations, we can propose the separation measure to be:
\begin{equation}
\Phi =\log \Big[1+2 \Big( \frac{\partial v(s)}{\partial s} \Big|_{s=\langle s_e \rangle}\Big) ^2 \frac{A}{\langle v(s_{e}) \rangle } \Big], 
\label{eqn:measure}
\end{equation}
where $A$ is the area underneath the auto-correlation function of the external noise:
\[
A = \lim_{t\rightarrow \infty}\int_0^\infty d \Delta t \Big[ \big\langle s_{e}(t+ \Delta t)s_{e}(t)\big\rangle - \langle s_{e}(t)\rangle^2 \Big].
\]
The area $A$ can be further simplified to be the variance of $s_e$ multiplied by its correlation time ($\tau$) as a first level of approximation.  The derivation of Eq.~\ref{eqn:measure} is given in the Supporting Material.  The measure increases with an increase in either  the sensitivity of the propensity function ($\partial v/\partial s$), the absolute strength of the external noise, or the correlation time of the external noise.  The measure, however, decreases with an increase in mean flux $\langle v(s_e) \rangle$ (Fig.~\ref{fig:crossover}).  This is counter-intuitive because: the larger the flux, the faster the concentration fluctuations and the wider the time scale separation.     However, the increase in the mean flux, depending on which parameters to control, can lead to an increase in the time scale separation via the change in the sensitivity.  E.g., If global proportionality constants are increased by $x\%$, both  the sensitivity and mean flux increase by $x \%$.  Thus, as a net effect, the measure can increase for this choice of control.

\subsection*{Power-law scaling in flux fluctuations}

We will now show briefly how the slope change is related to power law scaling that is observed in flow fluctuations in other complex networks \cite{Menezes2004, Menezes2004b,Eisler2005, Eisler2005b, Meloni2008}.   In the scaling studies, it was  investigated how the flux CV is related to mean flux (actually, rather than flux, but the number of events occurred within $\epsilon$, was investigated).   As shown in the Supporting Material, depending on the value of $\epsilon$ relative to the correlation time of the external noise, the scaling crossover takes different forms (Eqs.~S2 and S3).  We propose here that the scaling crossover that appears in other complex networks can also depend on the interplay between the external noise correlation time and $\epsilon$.    We note that only in the case of a \emph{linear} propensity function $v(s_e)= \alpha s_e$ for $\epsilon \ll \tau$,  we could regenerate the crossover function given in Eq.(7) in \cite{Meloni2008} (here the relative noise strength is given by $\mbox{Variance}(s_e)/\langle s_e \rangle^2$, and Eq.~S4 is used.)  We have therefore shown a connection between power-law scaling and flux fluctuations in reaction networks.  

\section*{SCA: Parametric Control of Noise Level}

In the literature on deterministic control theory \cite{Ingalls2006} and MCA \cite{Kacser1993,Small1994, Westerhoff1996, Kholodenko1998} some authors have considered the orthogonal control of  system variables such as flux and species concentrations. Here we consider the orthogonal control of mean concentration levels and concentration CVs, in order to control noise independently of the mean concentration levels.   Such control needs to satisfy the following requirements.  First, the concentration CV decreases as the concentration mean increases, and thus the control of mean and CV can be strongly anti-correlated. In this case one needs to find systematically which parameters to be perturbed by how much for the orthogonal control.  Second, the concentration CV is dependent on noise propagation \cite{Paulsson2004, Kim2008}, implying that a set of multiple parameters may need to be controlled simultaneously to achieve a sensitive change in CV.  Taking into account these requirements, we present a systematic non-local method for orthogonal control using the control coefficients.  

We introduce a control vector $C^y_{\bp} = (C^y_{p_1},C^y_{p_2}, \cdots, C^y_{p_L}) $ defined in an $L$-dimensional control parameter space.  When parameters $\bp$ are perturbed in the direction of $C^y_{\bp}$, a system variable $y$ (concentration mean or CV) shows a sensitive response of increase.  When $\bp$ are perturbed in one of the perpendicular directions of $C^y_{\bp}$, the system variable $y$ does not change.  

We aim to find parameter perturbations ($\blambda$) that lead to a decrease in the concentration CV without changing the concentration mean.  For the mean $\langle s \rangle $ not to be changed, the parameters must be perturbed in  the perpendicular directions of $C^{\langle s \rangle}_{\bp}$ (Fig.~\ref{fig:vector}).  It needs to be determined which one of the perpendicular directions leads to a sensitive response of a decrease in the concentration CV.  This determination can be done by projecting  $C^{V^{s}}_{\bp}$ onto the parameter space perpendicular to $C^{\langle s \rangle}_{\bp}$ and multiplying by $-1$:
\begin{equation}
{\blambda} \equiv (-1)\Big[ C^{\sigma^{s_n}}_{\bp} - \cos \theta |C^{\sigma^{s_n}}_{\bp}| \frac{C^{\langle s_n \rangle}_{\bp}}{|C^{\langle s_n \rangle}_{\bp}|}\Big],
\label{eqn:control-vector}
\end{equation}
where the factor of $-1$ appears since $V^s$ should decrease.

The efficiency of this orthogonal control can be estimated by how much percentage ratio of the control vector for CV is projected onto the perpendicular space: $| \sin \theta |$, where $\theta$ is the angle between the two control vectors.  If $\theta$ is close to $-180^\circ$, the two controls are anti-correlated and the efficiency is $\sim 0$.  If $\theta$ is close to $90^\circ$, the two controls are already orthogonal and the efficiency is $\sim 1$.    

We provide an example of orthogonal control to reduce the concentration CV by investigating a linear chain reaction system (Fig.~\ref{fig:linear}A).  This system is under negative feedback control and receives external noise via $S_1$.    We can predict the distribution of control based on the linear noise approximation (Fig.~\ref{fig:linear}B).  We have estimated control vectors for the concentration mean and CV, $C^{\langle s_4 \rangle}_{\bp}$ and  $C^{\sigma^{s_4}}_{\bp}$ (Fig.~\ref{fig:linear}B, crosses), by using the Lyapunov equation (refer to the Supporting Material) (also known as the fluctuation dissipation relationship \cite{Kubo1966, Paulsson2004}): 
\begin{equation}
\bJ \bsigma + \bsigma^T \bJ^T + \bD = 0,
\label{eqn:Lyapunov}
\end{equation}
with $\bJ$ the Jacobian matrix, $\bsigma$ concentration covariance matrix, and $\bD$ diffusion matrix.  We estimated $\theta$ and  ${\blambda}$ for the original parameter values. We perturbed the parameters along ${\blambda}$ and estimated the new $\theta$ and $\blambda$ for the perturbed ones.  After two more iterative perturbations,  we could reduce the noise level by 25\% without changing the mean level (Fig.~\ref{fig:linear}C).

\section*{SCA: Flux Fluctuation Control}

In this section, we discuss a way to reduce the flux CV.  Consider a scenario where a metabolic engineer aims to reduce the fluctuations in the production rate of an end product.  To this aim, her/his first guess is that reducing the concentration fluctuations will lead to a reduction in the rate fluctuations.  The engineer introduces a negative feedback to reduce the concentration fluctuations. The question we might ask is whether  this operation guarantees that the rate fluctuations are reduced?  

Let us consider the previous example of the linear-topology reaction system with negative feedback (Fig.~\ref{fig:linear}A).  We aim to reduce the fluctuations of $J_6$, by controlling $p_6$.   Based on Fig.~\ref{fig:linear}B, decreasing $p_6$ causes a reduction in the concentration CV of $S_4$.  We can decide to reduce $p_6$ to decrease the flux fluctuations.  To confirm that, we have estimated its flux control coefficients based on stochastic  simulations.  We  found that the sign of the control coefficients $C^{V^J_{66}}_{p_6}$ is however negative for $\epsilon \lesssim \tau_f$ ($\tau_f\sim 1$: feedback time scale) and positive for $\epsilon > \tau_f$.  Reduction of the concentration CV causes an increase in the flux CV for $\epsilon \lesssim \tau_f$, while it does not for $\epsilon > \tau_f$.   This means that controlling $p_6$ can have an opposite effect depending on $\epsilon$.  This is due to the fact that flux fluctuations become dominated by different sources of noise depending on $\epsilon$.   Therefore, in this case, we need to choose the appropriate value of $\epsilon$ depending on the rate fluctuations caused by which source of noise to be reduced.
   
A question that comes next is: why does the control distribution of flux CVs change with the value of $\epsilon$?   Consider again the negative feedback system.  There are three different time scales, related to the internal turn-over reactions ($\tau_i$), feedback controls ($\tau_f$), and external noise ($\tau \equiv 1/p_2$).  For our choice of parameter values, $\tau_i<\tau_f<\tau$.  Depending on where the value of $\epsilon$ resides, the flux CV control takes different distributions as shown in Fig.~\ref{fig:flux-dist}.  

For $\tau_f \lesssim \epsilon \lesssim \tau$,  the control distribution for downstream flux CVs is quite similar to that for the CV of $S_4$ (Figure \ref{fig:flux-dist}B ($\epsilon=40$) is compared with Fig.~\ref{fig:linear}B).   This is due to the strong negative feedback and the slow fluctuation components of $S_4$.  The reaction flux $J_3$ has been confirmed to be approximately equal to $v_3(S_1, S_4^*)$ with $S_4^*$ the slow component of fluctuations of $S_4$ (graph not shown).    

For $\epsilon \gtrsim \tau$, the external noise becomes averaged out.  Thus, the downstream flux CV becomes less sensitive to the external noise correlation time $\tau (= 1/p_2)$.  That is why the control by $p_2$ becomes weaker (see Fig.~\ref{fig:flux-dist}B($\epsilon=1000$)).   This change leads to the change in the sum value of control coefficients for downstream-flux CV, from $\sim 0$ to $\sim -1$.   

For $\epsilon \ll \tau_f$, the  internal noise becomes dominant.  All the downstream-flux CVs asymptotically follow $1/\epsilon \langle J \rangle$ with $\langle J \rangle$ the downstream-flux mean.  Therefore, the control vector for the downstream-flux CV is completely anti-parallel with that of $\langle J \rangle$ (Fig.~\ref{fig:flux-dist}B, $\epsilon =0.01$), implying that orthogonal control is impossible.

For $\epsilon \simeq \tau_f$, there is a hump in the plot of flux CV vs. $\epsilon$.  This is due to the negative feedback, where fluctuations of $S_4$ can be fed back at the same time scale without losing its control strength (see Fig.~\ref{fig:flux-dist}A).

\section*{Conclusion}
In this paper we describe extensions of metabolic control analysis into the stochastic regime for general biochemical reaction networks.  We have shown that there exist MCA-like summation theorems for stochastic sensitivity measures for mean values and coefficients of variation/covariation (CV/CCV) for concentrations and reaction fluxes.  The summation theorems for the reaction fluxes have shown that the sum values of control coefficients  for flux CVs/CCVs depend on the size of the measurement time window ($\epsilon$).  Such dependency becomes stronger as the reaction systems  shows multi-time-scale dynamics, i.e.  the time-scale separation between slow and fast modes becomes larger.  We have provided a measure to quantify such separation.      

In terms of the stochastic sensitivity measures, we have provided a non-local systematic way to control mean values and CVs of concentrations orthogonally.   We hope this method will be  useful for controlling noise levels in various reaction networks such as gene regulatory networks, metabolic reaction networks, and protein-protein interaction networks. 

Finally, we have shown that the control distribution of flux fluctuations can be significantly different depending on $\epsilon$ which reflects the dynamics at different time scales that emerge under varying values  of $\epsilon$.  Depending on which noise source the flux fluctuations to control is caused by, the appropriate window size needs to be chosen.

\vspace{0.2in}
This work was supported by a National Science Foundation (NSF) Grant in Theoretical Biology 0827592. Preliminary studies were supported by funds from NSF FIBR 0527023.  The authors acknowledge useful discussions with Hong Qian.



\clearpage

\begin{table}
\centering
\begin{tabular}{@{\vrule height 10.5pt depth4pt  width0pt}ll}
\hline\hline
$\langle f \rangle_{(x)}$ 	& Ensemble average of $f$ (over x) \vspace{-0.03in}\\
& at a stationary state\\
$p$&Control Parameter \\
$s$ & Concentration\\
$v$ & Reaction propensity function\\
$J$ & Reaction flux\\
$V_{ij}$ & Coefficient of co-variation (CCV) between $i$ and $j$\\
$V_{jj}$ & Coefficient of variation (CV) of $j$\\
\hline\hline
\end{tabular}
\caption{Notation}
\label{table-notation}
\end{table}

\clearpage
\section*{Figure Legends}
\subsubsection*{Figure~\ref{fig:2step}.}
Two step cascade reaction system:  $S_1$ down-regulates the reaction  creating $S_2$ (A).  The reaction rates involving $S_1$ are set 100 times slower than those involving $S_2$.  $S_1$ applies an external noise onto the (internal) system of $S_2$.  Time evolution of $S_1$ and $S_2$ is shown (B). The region of $t=[100, 120]$ is expanded (B,bottom).  The time evolution profile of $S_2$ follows the external noise with rapidly fluctuating internal noise (B,top).  In the time scale of the order of 1, $S_2$ does not fluctuate but $S_1$ fluctuates significantly, i.e., the internal noise becomes dominant (B,bottom).     $J_3$ is measured with three different time window sizes, $\epsilon = 0.0625,8,1024$ (C).  $J_3$ matches with $v_3$ for $\epsilon \simeq 8$, because the internal noise is averaged out, i.e., the external noise is dominant in this time scale (C,middle).   Flux variance of $J_3$ decreases with the time window size $\epsilon$ (C,D).   $V^J_{33}$ shows a plateau, while $V^J_{11}$ does not (D) ($V^J_{22}$ overlaps with $V^J_{11}$, and $V^J_{44}$ with $V^J_{33}$ [not shown in graph]).  The stochastic simulation algorithm \cite{Gillespie1977} was used.  Parameters: $(X_1,X_2,p_1,p_2,p_3,p_4) = (1,1,0.1,0.01,1,1)$. 

\subsubsection*{Figure~\ref{fig:2step-b}.}
Two step cascade reaction system (Fig.~\ref{fig:2step}A): The estimate of $V^J_{33}$ from the simulations is compared with the exact analytic result (Eq.~S4) and its asymptotic forms corresponding to $\epsilon \ll \tau(=1/p_2)$ and $\epsilon \gg \tau$.   The two asymptotic lines have slope -1 and their vertical separation (at the log-log scale) is denoted by $\Phi$, the time-scale separation measure (Eq.~\ref{eqn:measure}). 

\begin{figure}
\includegraphics[scale = 1]{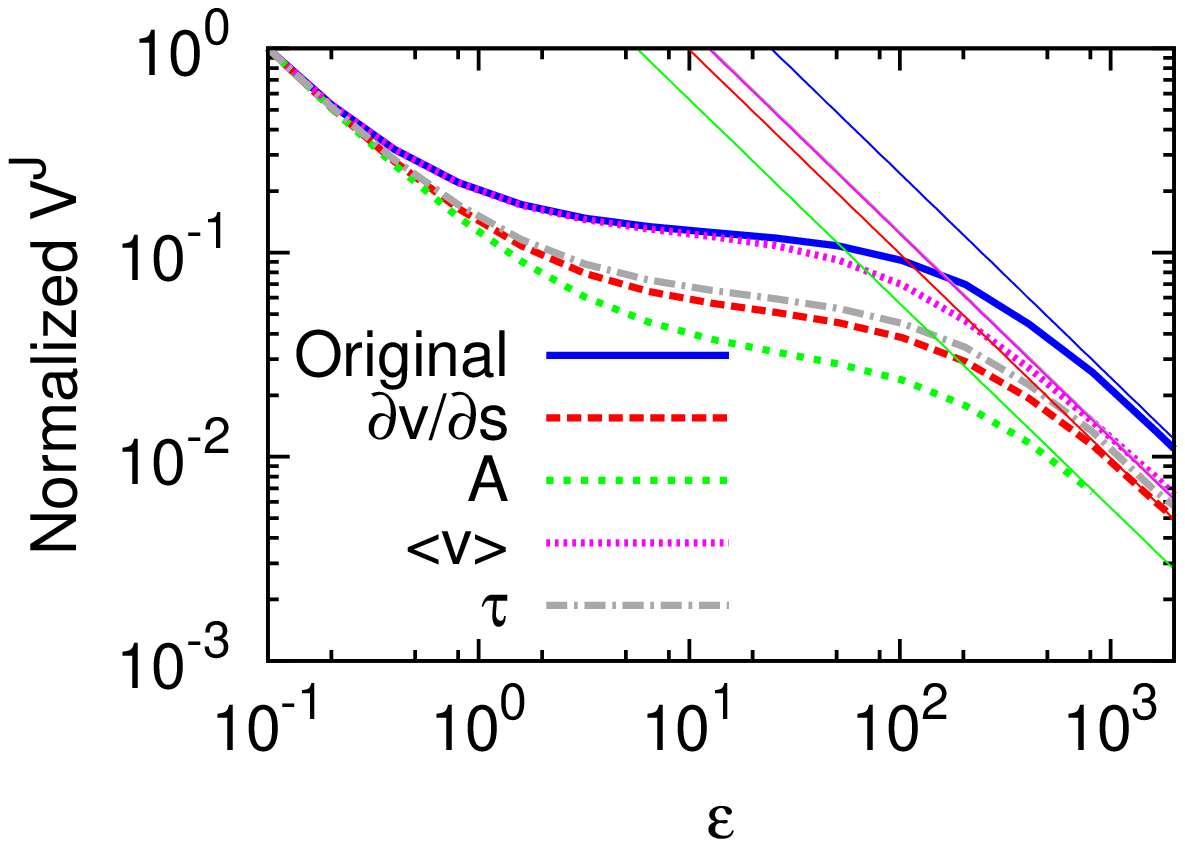}
\caption{}
\label{fig:crossover}
\end{figure}

\subsubsection*{Figure~\ref{fig:crossover}.}
Time-scale separation measure $\Phi$ (Eq.~\ref{eqn:measure}) is verified with numerical simulations.  External noise $s_e$, generated by $X_0 \xrightarrow{p_1X_0} S_e \xrightarrow{p_2 S_e} \O$,  is applied onto a reaction: $v(s_e) = p_3 +  \frac{p_4 s_e^n}{K_m + s_e^n}$.  The CV of the reaction flux of $v(s_e)$ is numerically estimated by using Eq.~S1 (solid line, Original).   We have reduced $\Phi$ by perturbing one or more factors affecting $\Phi$ for each case (other solid lines). We normalized $V^J$ such that its normalized value for $\epsilon=0.1$ equals 1 for  ease of comparison.  $\Phi$ given by Eq.~\ref{eqn:measure} is shown to predict the separation well (dotted lines: $\log(\mbox{Normalized $V^J$})=\Phi-1-\log_{10}(\epsilon)$).  Parameters used: $X_0 = 1$ for all the cases,  $(p_1,p_2,p_3,p_4,K_m,n) = (0.2,0.01,0,100,400,2)$ for ``Original", $(0.2,0.01,0,100,20,1)$ for ``$\partial v/\partial s$", $(0.4,0.01,0,100,400,2)$ for "A", $(0.2,0.01,100,100,400,2)$ for "$\langle v \rangle$", and $(0.4,0.02,0,100,400,2)$ for ``$\tau$".

\begin{figure}
\includegraphics[scale = 1]{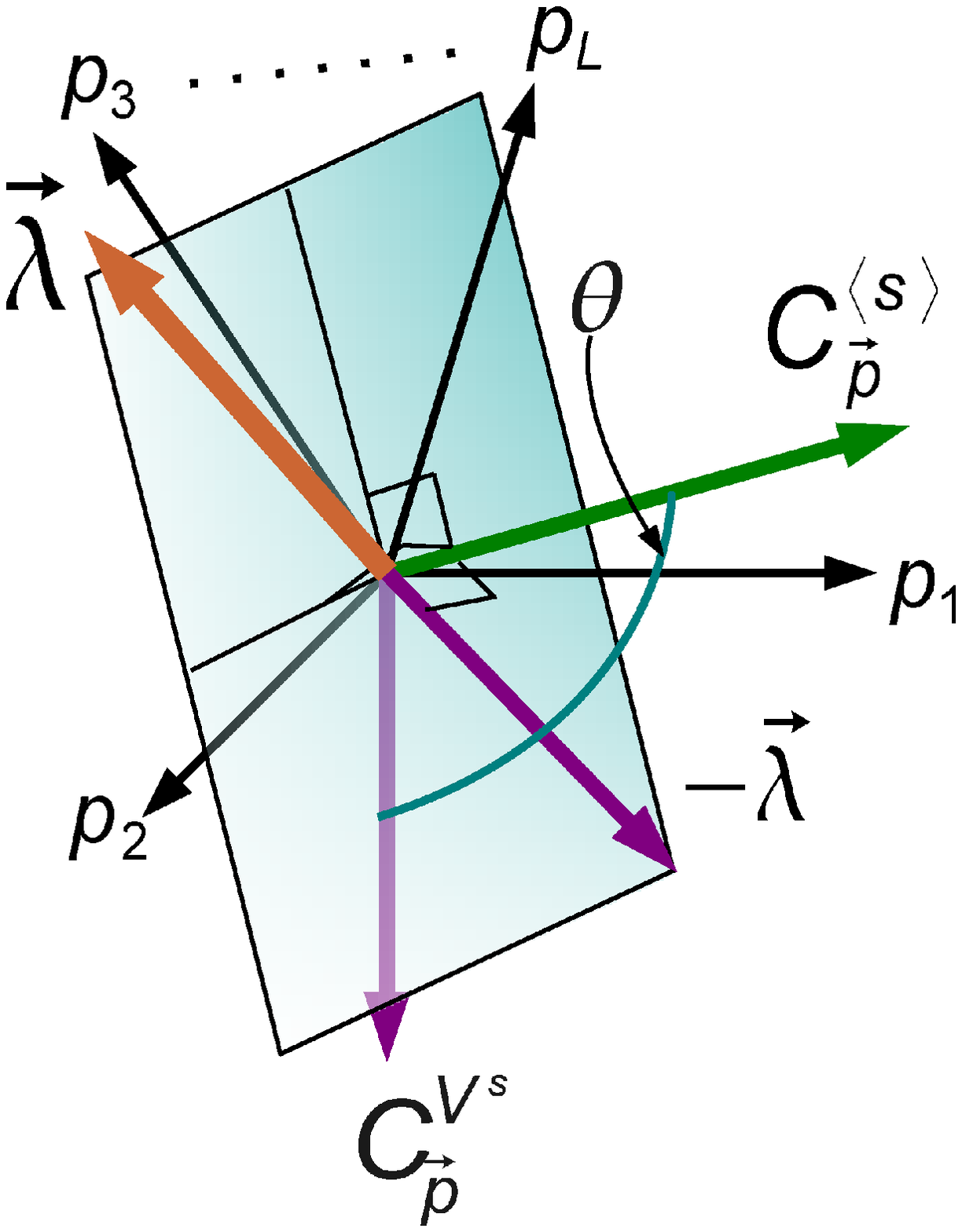}
\caption{}
\label{fig:vector}
\end{figure}

\subsubsection*{Figure~\ref{fig:vector}.}
Control vectors $C_{\bp}=(C_{p_1}, C_{p_2}, \cdots, C_{p_L})$  for a mean concentration, $\langle s \rangle$, and its CV, $V^s$,  are shown in an $L$-dimensional  parameter space as respective green and purple arrows, oriented in different directions ($\theta$: the angle in between).  The control vector $C_{\bp}^{V^s}$ is projected onto a space perpendicular to $C_{\bp}^{\langle s \rangle}$ (turquoise blue plane). The projected vector of $C_{\bp}^{V^s}$ is denoted by $-\blambda$.  When parameters perturbations are directed along $\blambda$, the CV shows a sensitive response of decrease while mean concentration remains the same.

\subsubsection*{Figure~\ref{fig:linear}.}
Orthogonal control of concentration CV for a linear topology reaction system with a negative feedback and under external noise (A).   Distributions of (scaled) control coefficients for mean values and CVs of all species are estimated by perturbing each parameter by 5\% (B) (simulation: hash and solid bars, linear noise approximation [shown only for $S_4$]: crosses).   The original parameter set is $(X_1, X_2, K_m)=(1,1,10^4), $ $\bp_0 = (p_1, p_2, p_3, p_4, p_5, p_6) =(0.1,0.01,2\times 10^4,1,1,1)$.  We aim to reduce  the noise level of $S_4$ without changing its mean.  Control vectors for $\langle S_4 \rangle$ and $V^S_{44}$ are estimated from the linear noise approximation.  Then, we estimated $\theta \sim 164^\circ$ and $\blambda \simeq (0.0, 0.0, -0.1, 0.1, 0.1, -0.1)$.  We perturbed  $\bp_0$ in the direction of $\blambda$ by 40\%, i.e., $\delta \bp_0  = (0,0, -8\times 10^3, 0.4,0.4,-0.4)$ and the new values of parameters $\bp_1$ are set to $\bp_0+\delta \bp_0 = (0.1,0.01,1.2\times 10^4, 1.4,1.4,0.6)$.  We repeated this procedure twice more.  The probability distribution functions of $S_4$ are shown for the series of the perturbations (C).  The final $\bp$ is $(0.1,0.01,5000,2.6,2.6,0.24)$. 
Efficiencies ($|\sin (\theta)|$) of orthogonal controls are shown for different values of a parameter $p_6$ for the other parameters fixed (D). Orthogonal control is most efficient around $0.2 \lesssim p_6 \lesssim 0.6$.

\subsubsection*{Figure~\ref{fig:flux-dist}.}
\begin{figure}
\vspace{-3cm}
\hspace{2cm}
\includegraphics[scale = 0.9]{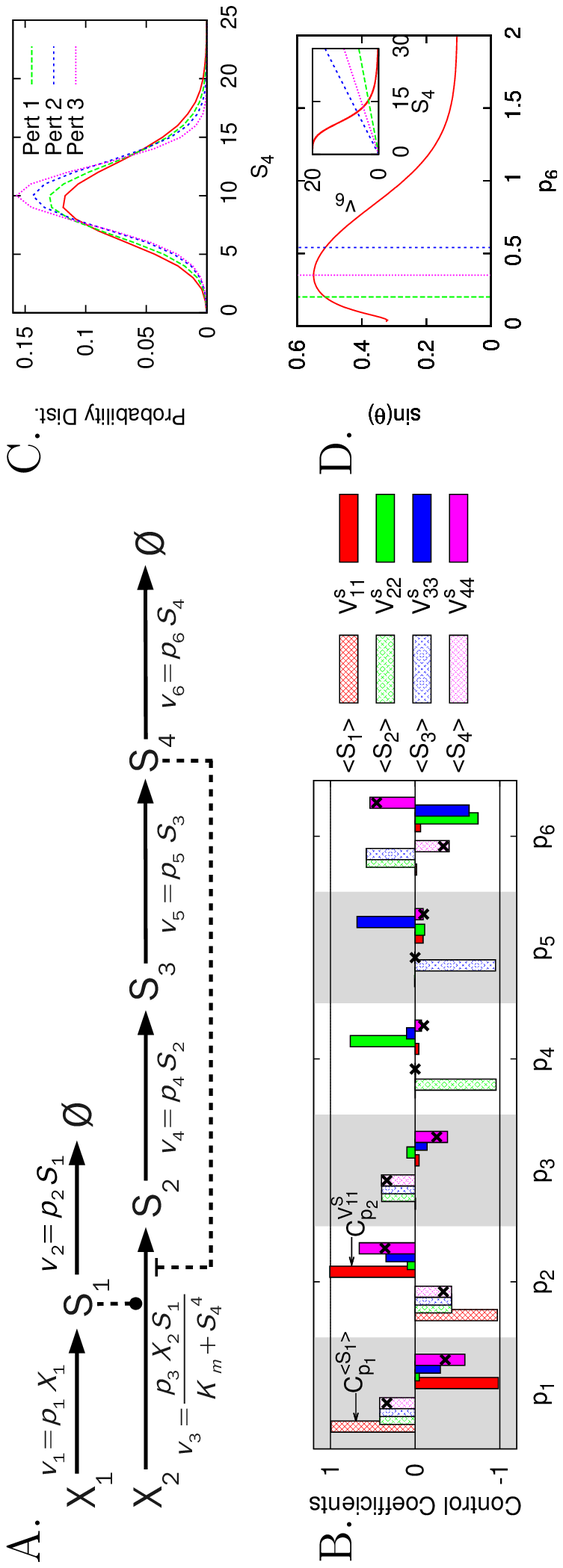}\\
\vspace{0cm}
\caption{}
\vspace{-7cm}
\label{fig:linear}
\end{figure}

\begin{figure}
\hspace{0cm}
\includegraphics[scale = 1]{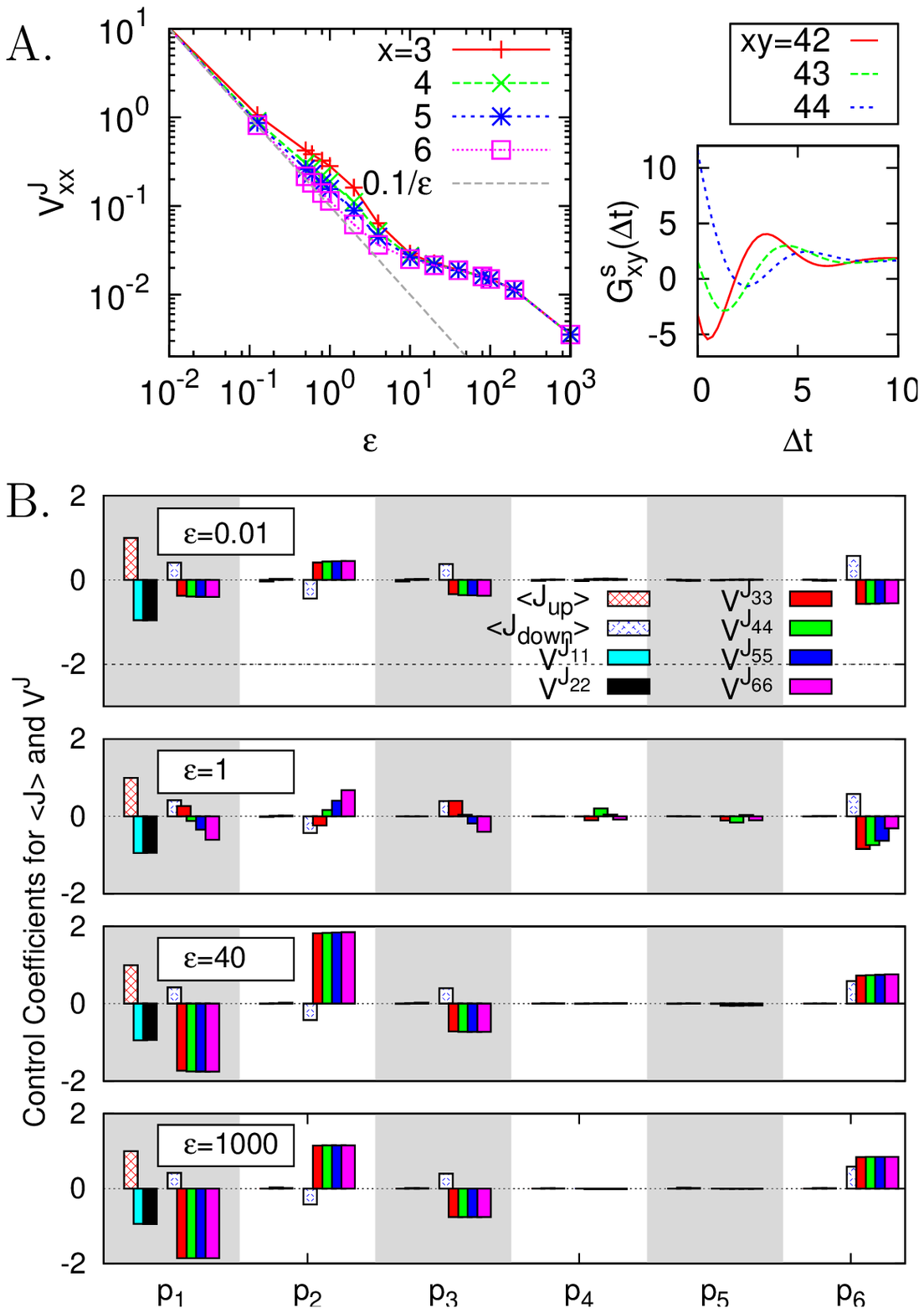}\\
\caption{}
\label{fig:flux-dist}
\end{figure}

Flux control distributions for the linear-topology reaction system with  negative feedback (Fig.~\ref{fig:linear}A). Flux CV of internal fluxes (A,left) shows humps in the time scale of the feedback ($\epsilon \sim 1$); correlation functions between $S_4$ and all internal species are shown on the right.  $G_{xy}^s(\Delta t) \equiv \lim_{t_0 \rightarrow \infty}\langle S_x (t_0+\Delta t)S_y(t_0) \rangle - \langle S_x(t_0) \rangle \langle S_y (t_0) \rangle$.  The parameter set $\bp_0$ is used (Fig.~\ref{fig:linear} caption.) Control distributions (control vectors) for flux CVs are significantly dependent on $\epsilon$. For $\epsilon=0.01$, the control vector for each mean flux is anti-parallel with that for each corresponding flux CV. For $\epsilon=40$, control distribution becomes similar to the control distributions of CV of $S_4$ (Fig.~\ref{fig:linear}B.)

\clearpage

\clearpage

\clearpage

\clearpage

\clearpage

\clearpage

\clearpage
\begin{center}
\Large{Stochastic Control Analysis for Biochemical Reaction Systems\\ (Supporting Material)}
\end{center}

\begin{center}
\large{{Kyung~Hyuk~Kim and Herbert~M.~Sauro}}
\end{center}

\section{Time scale separation measure}
We consider a reaction step  with its propensity function given by $v(s_{e})$,  showing a dependence on the external noise.  If events in $v(s_e)$ do not cause large fluctuations of its substrate concentrations, the counting process of the reaction events can be described by a doubly-stochastic Poisson process \cite{Cox1955}.  The probability $P(n;\epsilon)$ of having the number $n$ of events of reaction $v$ within the time window $\epsilon$ is given by 
\[
P(n,\epsilon) = \frac{1}{n!}  \sum_{\{s_e(t)\}} P(\{s_e(t)\})  N(\epsilon)^n e^{-N(\epsilon)},
\]
where $P(\{s_e(t)\})$ is the probability of having a trajectory of $\{ s_e(t)\}$ for the region  $t \in [0,\epsilon]$,  and $N(\epsilon) \equiv  \int_0^\epsilon dt \: v(s_e(t))$.  The CV of its flux is given by \cite{Cox1955}
\begin{equation}
V^J = \frac{1}{\epsilon \langle J \rangle}\Big(  1 + \frac{\mbox{Variance}(N(\epsilon))}{\epsilon \langle J \rangle} \Big),
\label{eqn:VJ}
\end{equation}
where 
\[
\mbox{Variance}(N(\epsilon)) = \int_0^\epsilon  dt_1 \int_0^\epsilon dt_2  \big\langle \delta v(s_e(t_1)) \delta v(s_e(t_2)) \big\rangle_{s_e},
\]
with $\delta v(s_e) \equiv v(s_e) - \langle v(s_e) \rangle_{s_e}$. \vspace{0.1in}

For $\epsilon \ll  \tau$, $\mbox{Variance}(N(\epsilon))$ can be simplified as 
\[
\mbox{Variance}(N(\epsilon)) \simeq \epsilon^2 \Big\langle   \big[ \delta v(s_e(t))\big]^2 \Big\rangle.
\]
Thus, we obtain
\begin{equation}
V^J \simeq \frac{1}{\epsilon \langle J \rangle}\Big(1 + \epsilon \frac{\mbox{Variance}(v(s_e))}{\langle J \rangle} \Big), \,\,\,\,\,\mbox{for}\,\,\,\, \epsilon \ll \tau.
\label{eqn:crossover1}
\end{equation}

For $\epsilon \gg \tau$, $\mbox{Variance}(N(\epsilon))$ is simplified as
\[
\mbox{Variance}(N(\epsilon)) \simeq 2 \int_0^\epsilon dt_1 \int_0^{\infty} dt' G_{\delta v(s_e)} (t'),
\]
where $G_v(t')$ denotes an autocorrelation function defined by $\langle \delta v(t_0+t')\delta v(t_0) \rangle_{s_e}$.  This can be further  simplified by $2 \epsilon A'$, with $A' \equiv \int_0^\infty dt' G_{\delta v(s_e)} (t')$.   $A'$ is the area underneath the autocorrelation function.  Thus, we obtain
\begin{equation}
V^J \simeq \frac{1}{\epsilon \langle J \rangle} \Big(1 + \frac{2A'}{\langle J \rangle}\Big),\,\,\,\,\,\mbox{for}\,\,\,\, \epsilon \gg \tau
\label{eqn:crossover2}
\end{equation}

If the fluctuations in $s_e$ is mostly confined to the linear region of $v(s_e)$, then $\delta v(s_e) \simeq \alpha  \delta s_e$ with $\alpha \equiv \frac{\partial v(s)}{\partial s}|_{s=\langle s_e \rangle}$.  Thus, we obtain two different asymptotic forms of the flux CV for $\epsilon \ll \tau$ and $\epsilon \gg \tau$, respectively:
\begin{equation}
V^J = \frac{1}{\epsilon \langle J \rangle} \Big[  1+  \epsilon \Big(\frac{\partial v}{\partial s_e} \Big)^2 \frac{\mbox{Variance}(s_e)}{\langle J \rangle} \Big], \,\,\,\,\,\mbox{for}\,\,\,\, \epsilon \ll \tau,
\label{eqn:measure-small}
\end{equation}
and
\begin{equation}
V^J = \frac{1}{\epsilon \langle J \rangle} \Big[  1+  2 \Big(\frac{\partial v}{\partial s_e} \Big)^2 \frac{A}{\langle J \rangle} \Big], \,\,\,\,\,\mbox{for}\,\,\,\, \epsilon \gg \tau,
\label{eqn:measure-large}
\end{equation}
where $A \equiv \int_0^\infty dt' G_{\delta s_e}(t')$ is the area underneath of the autocorrelation function of the concentration of the external species, $s_e$.

The time scale separation measure is defined as the vertical distance between the two asymptotic linear lines for the log-log plot of flux CV vs. $\epsilon$ corresponding to $\epsilon \rightarrow 0$ and $\infty$.  The measure is obtained from Eq.\eqref{eqn:measure-small} and \eqref{eqn:measure-large}:
\[
\Phi = \log \Big[  1+  2 \Big(\frac{\partial v}{\partial x} \Big)^2 \frac{A}{\langle J \rangle} \Big].
\]

For the two-step reaction process as shown in Fig.1A,  we can obtain the following exact result from Eq.\eqref{eqn:VJ} without any approximation: 
\begin{equation}
V^J_{33} =\frac{1}{\epsilon \langle J_3 \rangle}+ \frac{2}{\langle S_1 \rangle} \frac{\chi -1 + e^{-\chi}}{\chi^2},
\label{eqn:exact}
\end{equation}
where $\chi \equiv p_2 \epsilon$. The second term  converges to $1/2$ for $\chi \ll 1$  and vanishes as $1/\chi$ for 
$\chi \gg 1$. While control parameters are fixed, we vary the value of $\epsilon$ (see Fig.2).  $V^J_{33} \simeq 1/\epsilon \langle J_3 \rangle$ for $\epsilon \ll 1/p_2$.  As  $\epsilon$ increases, the  flux variance reaches a plateau region following $1 / \epsilon \langle J_3 \rangle + 1/\langle S_1 \rangle$.   As $\epsilon \gg  1/p_2$, it follows $(1/\langle J_3 \rangle+ 2/\langle S_1 \rangle p_2 ) / \epsilon$.  The time scale separation measure for this system is expressed as 
\[
\Phi = \log \Big[1+ \frac{2\langle J_3 \rangle }{\langle J_1 \rangle}\Big].
\]
The measure increases with the internal flux $\langle J_3 \rangle$ and  the time scale separation becomes larger as the internal dynamics becomes faster.

\section{Coefficients of covariance of fluxes vs. $\epsilon$} \hspace{2mm}
In this section, we will investigate how the sum value of Eq.[4] changes with $\epsilon$ for the coefficients of covariation(CCV) between two different fluxes by investigating the slope of the log-log plot of flux CCV vs. $\epsilon$.     For ease of presentation, we will consider covariances of fluxes rather than the coefficients of covariation.   A covariance between two different fluxes is defined as 
\[
\sigma^J_{ij} = \Big\langle \big(J_i-\langle J_i\rangle\big)\big( J_j-\langle J_j\rangle\big) \Big\rangle = \langle J_i J_j \rangle - \langle J_i \rangle \langle J_j \rangle,
\]
where the ensemble average $\langle . \rangle$ is performed over the stationary states obtained by independent runs of stochastic simulations. 

\begin{figure}
\begin{center}
\includegraphics[scale = 0.5, angle = 0]{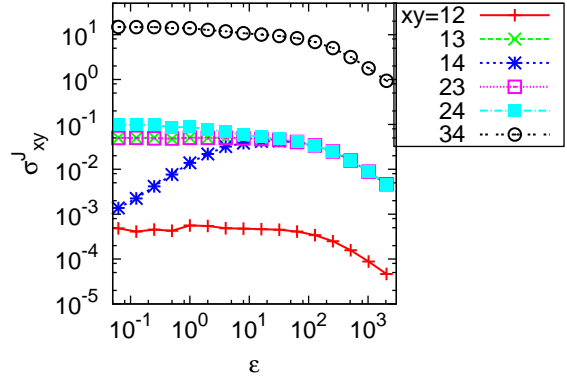}
\end{center}
\caption{Flux covariances of different pairs of reactions in the two step cascade reaction system Fig.1A.  Parameters: $(X_1,X_2,p_1,p_2,p_3,p_4) = (1,1,0.1,0.01,1,1)$.}
\label{fig:2step-J-cov-si}
\end{figure}

Consider a two-step cascade reaction system as shown in Fig.1A.  First, we will investigate how the flux covariance behaves in the limit of $\epsilon \rightarrow 0$.  Flux covariances show different asymptotic behaviors in the limit of $\epsilon \rightarrow 0$ depending on the different pairs of fluxes (see Fig.\ref{fig:2step-J-cov-si}).  We will explain the mechanisms that generate the different  behaviors. 

\begin{figure}
\begin{center}
\includegraphics[scale = 0.5, angle = 0]{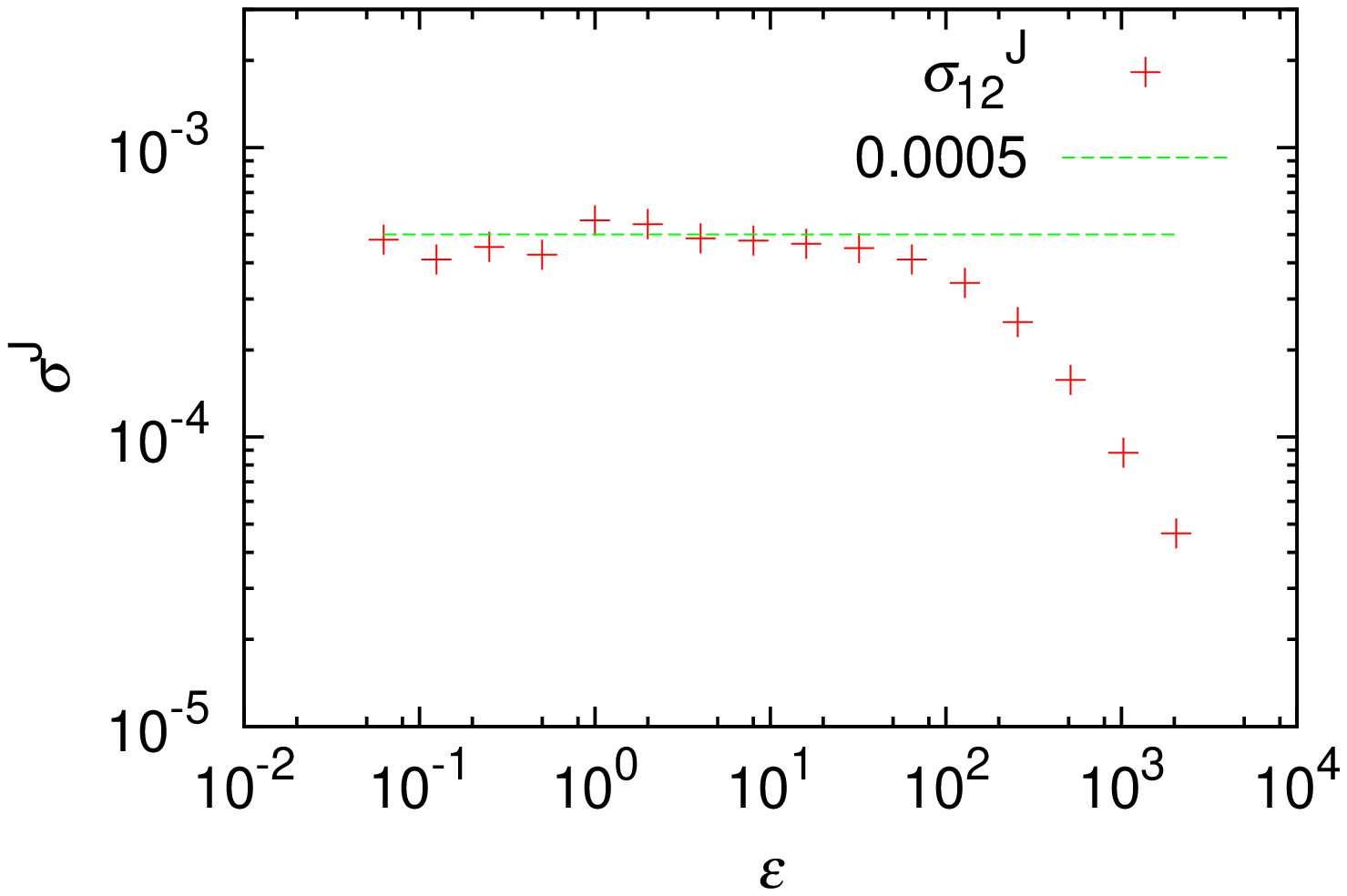}
\end{center}
\caption{Flux covariance of two reactions $v_1$ and $v_2$ in the two step cascade reaction system Fig.1A.  Parameters: $(X_1,X_2,p_1,p_2,p_3,p_4) = (1,1,0.1,0.01,1,1)$.}
\label{fig:2step-J-cov-1-2}
\end{figure}

First, we investigate the flux covariance between $J_1$ and $J_2$.  If we assume that $J_1$ and $J_2$ become independent in the limit of  $\epsilon \rightarrow 0$, the covariance $\sigma^J_{12}$ vanishes.  This, however, is not what we observed by simulation.   This indicates that there is a correlation between them.  The correlation is due to the fact that one reaction of  $v_1$ will increase $S_1$ by one, resulting in the increase of $v_2$ and affecting the proabability that the reaction $v_2$ will occur.   We take into account this \emph{causal} correlation to estiamte the flux covariance.  For a sufficiently small value of $\epsilon$,  the dominant contributions to the flux covariance come from  two cases: first,  reactions of $v_1$ and $v_2$ occur once for each within the time interval $\epsilon$, with the reaction $v_1$ first and then the reaction $v_2$, and second, each reaction occurs in the opposite order.    The contribution of the first case to the estimation of $\langle J_1 J_2 \rangle$   is, for sufficiently small $\epsilon$,
\[
\frac{1}{\epsilon^2} \int_0^\epsilon dt \int_t^\epsilon dt' v_1 v_2 (S_1+1) P_{s}(S_1, S_2) \simeq \frac{\big\langle v_1 v_2(S_1+1)\big\rangle }{2}.
\]
The contribution of the second case is 
\[
\frac{\big\langle v_1 v_2(S_1)\big\rangle}{2}.
\]
Thus, we obtain the covariance:
\[
\sigma^J_{12} \simeq \frac{\langle v_1 v_2(S_1+1) + v_1 v_2(S_1)\rangle }{2} - \langle v_1 \rangle \langle v_2 \rangle.
\]
Since $v_1$ is constant ($v_1 = p_1$) and $v_2  = p_2 S_1$, we obtain
\[
\sigma^J_{12} \simeq  \frac{1}{2}v_1 p_2 = \frac{1}{2}p_1 X_1 p_2.
\]
We have verified this result with the simulation data as shown in Fig.~\ref{fig:2step-J-cov-1-2}.

\begin{figure}
\begin{center}
\includegraphics[scale = 0.5, angle = 0]{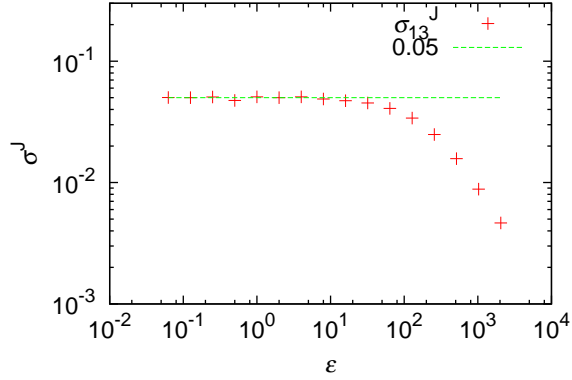}
\end{center}
\caption{Flux covariance of two reactions $v_1$ and $v_3$ in the two step cascade reaction system Fig.1A.  Parameters: $(X_1,X_2,p_1,p_2,p_3,p_4) = (1,1,0.1,0.01,1,1)$.}
\label{fig:2step-J-cov-1-3}
\end{figure}

The flux covariance between $J_1$ and $J_3$ in the limit of $\epsilon\rightarrow 0$ can be also estimated in the same way as above:
\[
\sigma^J_{13} \simeq  \frac{1}{2}v_1 p_3 = \frac{1}{2}p_1 X_1 p_3.
\]
The covariance is estimated at 0.05 (see Fig.\ref{fig:2step-J-cov-1-3}).

\begin{figure}
\begin{center}
\includegraphics[scale = 0.5, angle = 0]{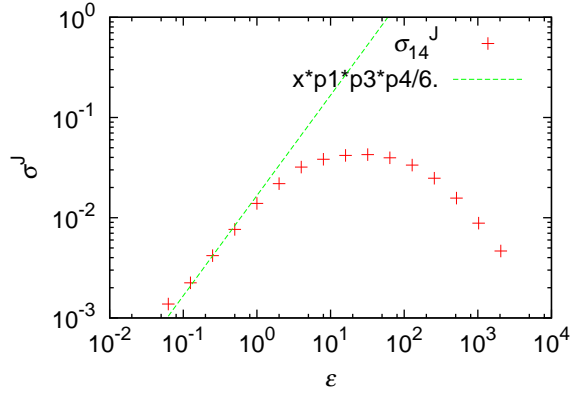}
\end{center}
\caption{Flux covariance of two reactions $v_1$ and $v_4$ in the two step cascade reaction system Fig.1A.  Parameters: $(X_1,X_2,p_1,p_2,p_3,p_4) = (1,1,0.1,0.01,1,1)$.}
\label{fig:2step-J-cov-1-4}
\end{figure}

$\sigma^J_{14}$ converges to 0 linearly with $\epsilon$ as $\epsilon \rightarrow  0$.   This is because an event of reaction $v_1$ does not make any change in the number of $S_2$.  The only way to make a correlation between $J_1$ and $J_4$ is through an event of reaction $v_3$.  By taking into account such indirect effects, the contribution to $\langle J_1 J_4\rangle$ becomes
\begin{eqnarray*}
\lefteqn{ \hspace{-0.6in}\frac{1}{\epsilon^2}\int^\epsilon_0 dt \int^\epsilon_t dt' \int^\epsilon_{t'} dt'' v_1 v_3(S_1+1)v_4(S_2+1) P(S_1, S_2)}\\
&=& \frac{1}{6}p_1 p_3 p_4 \big\langle(S_1+1)(S_2+1)\big\rangle \epsilon.
\end{eqnarray*}
Since the non-zero effect on $\sigma^J_{14}$ comes from the three-event correlation,  we obtain 
\[
\sigma^J_{14} = \frac{1}{6}p_1 p_3 p_4  \epsilon,
\]
and this result is verified with the simulation data as shown in Fig.~\ref{fig:2step-J-cov-1-4}.

The covariance between $J_2$ and $J_3$ shows a plateau region for the small value of $\epsilon \lesssim 1$ and this occurance is  due to the fact that $J_2$ and $J_3$ are causally  correlated and also that they share a common source of noise.  $\langle J_2 J_3 \rangle$ are estimated by considering two cases of event sequences: one event of $v_1$ comes first and then $v_2$ later, and these events occur in the opposite order. By taking into account both the cases, we can estimate $\langle J_2 J_3 \rangle$ as
\[
\langle J_2 J_3 \rangle  = \frac{1}{2}\langle v_2(s_1) v_3 (s_1-1) \rangle + \frac{1}{2} \langle v_3 (s_1) v_2 (s_1) \rangle,
\]
where the first term represents the case that an event of reaction $v_2$ occurs first, resulting in the decrease in $S_1$ by one, and then an event of reaction $v_3$ occurs.  The second term is for the other case that the reactions occur in the opposite order.  Therefore, we obtain the flux covariance:
\[
\sigma^J_{23} \simeq p_2 p_3 \big[ \sigma^s_{11}- \frac{1}{2}\langle S_1 \rangle  \big].
\]   
The first term on the left hand side is due to the common source of noise, in this case $S_1$, and the second due to the causal correlation.  The above expression can be further simplified to    $\sigma^J_{23} \simeq  \frac{1}{2}p_2 p_3 \langle S_1 \rangle$.     The height of the plateau is well estimated at $0.05$ (graph is not shown).

The covariance between $J_2$ and $J_4$ also shows a plateau region for the small value of $\epsilon \lesssim 1$, and the height of the plateau can be estimated by 
\[
\sigma^J_{24} = \langle v_2(S_1) v_4(S_2) \rangle  - \langle v_2 \rangle \langle v_4 \rangle = p_2 p_4 \sigma^s_{12}.
\]
This estimates the plateau height well (graph is not shown).
The reason for the occurance of the plateau region is that $J_2$ and $J_4$ have a common source of noise, resulting in the flux covariance: E.g., an event of reaction $v_2$ can be correlated with that of reaction $v_4$ by events of reaction $v_1$ that has occurred previously.

\begin{figure}
\begin{center}
\includegraphics[scale = 0.5, angle = 0]{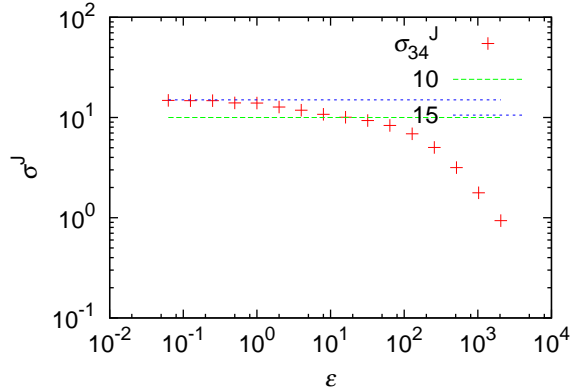}
\end{center}
\caption{Flux covariance of two reactions $v_3$ and $v_4$ in the two step cascade reaction system Fig.1A.  Parameters: $(X_1,X_2,p_1,p_2,p_3,p_4) = (1,1,0.1,0.01,1,1)$.}
\label{fig:2step-J-cov-3-4}
\end{figure}

$\sigma^J_{34}$ can be estimated by following the simliar estimation procedure to the one for $\sigma^J_{23}$:
\[
\sigma^J_{34} = p_3 p_4 \big[ \sigma^s_{12}+ \frac{1}{2}\langle S_1 \rangle   \big]
\]
The first term is due to the noise propagation from common sources of noise and the second due to the causal correlation. The flux covariance is estimated at 15 (see Fig.\ref{fig:2step-J-cov-3-4}).  

Finally, for the intermediate and  large value of $\epsilon$, i.e.,  $\epsilon \gtrsim 50$, four different covariance quantities match with one another: $\sigma^J_{13}$, $\sigma^J_{23}$, $\sigma^J_{24}$, $\sigma^J_{14}$; $J_1 \simeq J_2$ and $J_3 \simeq J_4$.

In summary, the sum value of the flux CV summation theorem depends on which reaction pairs to choose as well as the value of $\epsilon$.  The asymptotic forms of flux CCVs in the limit of $\epsilon \rightarrow 0$ are independent of $\epsilon$, i.e., plateau regions appear, if (1) the two reaction steps are affected by the noise propagated from common sources or (2) they are directly connected such that one reaction event leads to the direct change in the probability that the other reaction occurs.    

\section{Estimation of control coefficients for concentration CVs from the Lyapunov equation}
In this section, we will  show how we estimate control coefficients for concentration CV based on the linear noise approximation.   Let us define a mathematical notation:  The matrix component $(i,j)$ of $\frac{ \partial \by}{\partial \bx}  $ is $\frac{ \partial y_i}{\partial x_j}$.  $\frac{\delta x}{\delta p_i}$ denotes the change in the system variable $x$ from one stationary state to another due to a parameter perturbation of $p_i \rightarrow p_i+ \delta p_i$. 

Consider an infinitesimal perturbation in the control parameters denoted by $\bp$.  The Lyapunov equation Eq.[7] \cite{Kubo1966,Paulsson2004} is invariant because we consider stationary state perturbations:
\[
\delta( \bJ \bsigma + \bsigma^T \bJ^T + \bD) =0.
\]
   We obtain 
\begin{equation}
\bJ \frac{\delta \bsigma}{\delta p_i} +\frac{\delta \bsigma}{\delta p_i}\bJ^T + \frac{\delta \bJ}{\delta p_i} \bsigma + \bsigma \frac{\delta \bJ^T}{\delta p_i} + \frac{\delta \bD}{\delta p_i}=0, 
\label{eqn:unscaled}
\end{equation}
where we have used $\bsigma= \bsigma^T$ and $\delta \bsigma/\delta p_i$ means the change in the concentration covariance matrix due to the change in $p_i$ and it is defined as an unscaled control coefficient of the concentration covariance matrix, $\bsigma$.  This unscaled control coefficient will be estimated first and then the scaled control coefficient of concentration CV will be obtained later. 

To solve the above equation for $\delta \bsigma/\delta p_i$, we need to express $\delta \bJ / \delta p_i$ and $\delta \bD/ \delta p_i$ in terms of concentrations $\bs$ and $\bp$.   $\delta \bJ (\bs, \bp) / \delta p_i$ can be expressed as follows:
\[
\frac{\delta \bJ(\bs,\bp)}{\delta p_i}=\frac{\partial \bJ}{\partial p_i} + \frac{\partial \bJ}{\partial \bs} \frac{\partial \bs}{\partial p_i}.
\]
 By performing the similar procedure for $\bD$, $\delta \bD/\delta p_i$ can be expressed as:
\[
\frac{\delta \bD(\bs,\bp)}{\delta p_i}=\frac{\partial \bD}{\partial p_i} + \frac{\partial \bD}{\partial \bs} \frac{\partial \bs}{\partial p_i}.
\]
By substituting the above two expressions in Eq.\eqref{eqn:unscaled}, the unscaled control coefficients for a concentration covariance matrix ($\delta \bsigma/\delta p_i$) can be numerically estimated. 

Next, we need to obtain the control coefficients for concentration CV/CCV instead of concentration variance/covariance.  The concentration CV is defined as $V^s_{jk} = \sigma_{jk}/ s_j s_k$.   The unscaled control coefficients for the concentration CV can be obtained:
\begin{equation}
\frac{\delta V_{jk}}{\delta p_i} = \frac{1}{s_j s_k} \frac{\delta \sigma_{jk}}{\delta p_i}  - \frac{\sigma_{jk}}{s_j^2 s_k}\frac{\delta s_j}{\delta p_i} - \frac{\sigma_{jk}}{s_j s_k^2}\frac{\delta s_k}{\delta p_i},
\label{eqn:unscaled_V}
\end{equation}
where $\delta s_j/\delta p_i$ is an unscaled control coefficient for mean concentration $s_j$.   In this section, we have obtained the mathematical forms of control coefficients, under the assumption of the linear noise approximation.  At this approximation level, the mean concentration dynamics are described by the deterministic rate laws, by neglecting the contributions of all concentration covariances and higher moments \cite{Kim2008}.  Thus, we can express the unscaled concentration control coefficients as in the deterministic case:
\[
\frac{\delta \bs}{\delta p_i} = -\bJ^{-1} \bN_R \frac{\partial \bv}{\partial p_i},
\]
where $\bN_R$ is a reduced stoichiometry matrix \cite{Reder1988}.  
By substituting both this expression for $\delta \bs / \delta p_i$ and the numerical estimate of $\delta \bsigma/\delta p_i$ in Eq.\eqref{eqn:unscaled_V}, the unscaled control coefficients for the concentration CV/CCV can be estimated.  

Finally, we convert the unscaled control coefficient to a scaled version by using:
\[
C^{V^s_{jk}}_{p_i } = \frac{p_i}{V^s_{jk}} \frac{\delta V^s_{jk}}{\delta p_i}.
\]

We provide a MATHEMATICA file for this estimation in the Supporting MATHEMATICA file.

\end{document}